\documentclass[twocolumn,showpacs,preprintnumbers,amsmath,amssymb,letterpaper]{revtex4}
\usepackage{graphicx}% Include figure files
\usepackage{dcolumn}% Align table columns on decimal point
\usepackage{bm}% bold math
\usepackage{amssymb}
\usepackage{epsfig}
\newcommand{\czval}{0.1955}
\newcommand{\czerr}{0.0011}
\newcommand{\coval}{-0.1943 }
\newcommand{\coerr}{0.0044 }
\newcommand{\bttval}{-0.3573 }
\newcommand{\btval}{-0.29315 }
\newcommand{\bterr}{ 0.0005}
\newcommand{\btabs}{-0.2930}
\newcommand{\btabspercent}{0.7\%}
\newcommand{\btterr}{ 0.0005}
\newcommand{\discrep}{ 0.0021}
\newcommand{\be}{\begin{equation}}
\newcommand{\ee}{\end{equation}}
\newcommand\beq{\begin{eqnarray}}
\newcommand\eeq{\end{eqnarray}} 
\newcommand\eqn[1]{\label{eq:#1}} 
\newcommand\eq[1]{eq. (\ref{eq:#1})} 
\newcommand\eqstwo[2]{eqs. (\ref{eq:#1},\ref{eq:#2})}

\newcommand{\bfk}{{\mathbf k}}
\newcommand{\bfq}{{\mathbf q}}
\newcommand{\bfp}{{\mathbf p}}
\newcommand{\bfP}{{\mathbf P}}

\newcommand{\CD}{{\cal D}}

\newcommand{\CG}{{\cal G}}

\newcommand{\CT}{{\cal T}}

\newcommand{\CL}{{\cal L}}

\newcommand{\Tr}{{\rm Tr\,}}
\newcommand\half{{\textstyle{\frac{1}{2}}}}

\newcommand{\mybar}[1]%
        {\kern 0.6pt\overline{\kern -0.6pt#1\kern -0.6pt}\kern 0.6pt}
\begin{document}

\preprint{INT-PUB-11-017}

\title{A new field theoretic method for the virial expansion}

\author{David B. Kaplan}
\email{dbkaplan@uw.edu}

\author{Sichun Sun}
 \email{sichun@uw.edu}
 
 \affiliation{Institute for Nuclear Theory, Box 351550, Seattle, WA 98195-1550, USA}
 
 \date{\today}
 
 \begin{abstract}
We develop a graphical method for computing the virial expansion coefficients for a nonrelativistic quantum field theory.  As an example we compute the third virial coefficient $b_3$ for unitary fermions, a nonperturbative system.  By calculating several graphs and performing an extrapolation, we arrive at $b_3 =\btabs$, within $\btabspercent$ of a recent computation  $b_3 = -0.29095295$ by Liu, Hu and Drummond  \cite{liu2009virial}, which involved summing 10,000 energy levels for three unitary fermions in a harmonic trap.

\end{abstract}

\maketitle

\section{Introduction}

The virial expansion  allows one to  express the equation of state of a non-ideal gas in a density expansion, and is equivalent to a fugacity expansion of the grand potential density at nonzero chemical potential:
\beq
-\frac{\beta\Omega}{V} = \beta P =\frac{2}{\lambda^3}\left[z+ b_2 z^2 + b_3 z^3 + \ldots\right]
\eeq
or $\partial\Omega/\partial\mu = -2(V/\lambda^3) \sum_n n\, b_n z^n$. Here $V$, $P$ and $\beta$ are the volume, pressure and inverse temperature respectively, $z=e^{\beta\mu}$ is the fugacity, and $\lambda = \sqrt{2\pi\beta/M}$ is the thermal wavelength; the $b_n$ are dimensionless quantities directly related to the  virial coefficients. The $O(z)$ contribution is independent of interactions, and therefore the ideal gas term $2/\lambda^3$ has been factored out front (we assume here a gas of spin 1/2 fermions).  The thermal wavelength  provides a natural length scale, and the fugacity expansion is expected to be valid when $\lambda$ is short compared to the average inter-particle distance, and long compared to the range of interactions.  This expansion is of  current interest because of experimental focus on the properties of dilute atomic gases; and the case of fermionic atoms at a Feshbach resonance --- where the two-body scattering length diverges --- is of particular interest to both theorists and experimentalists.  Such examples of ``unitary fermions"  are  strongly interacting conformal  systems interpolating between the BCS and BEC regimes, with universal properties that serve (on a completely different length scale) as an interesting starting point for effective field theory treatments of interacting nucleons  \cite{Kaplan:1998tg,Kaplan:1998we}.   There has been extensive theoretical interest in computing  the parameter $b_3$ in the fugacity expansion for unitary fermions  \cite{Bedaque:2002xy,rupak2007universality}, culminating in a high accuracy determination $b_3=-0.29095295$ from a spectral study of the 3-fermion system, solving for the lowest 10,000 energy levels for three unitary fermions in a harmonic trap  \cite{liu2009virial}, a value that appears to agree with experimental results  \cite{PhysRevLett.98.080402,2010Natur.463.1057N}.

It would be  convenient to have a method for computing the $b_n$ coefficients directly using field theory techniques, particularly if accurate results could be obtained by computing a small set of diagrams.  Despite the fact that $\Omega$ is directly related to the sum of one particle irreducible Feynman diagrams in a finite temperature field theory,  the theory is not ideally suited to this task for several reasons: interparticle potentials are not in general readily described by Feynman rules; the sum of multiple interparticle interactions cannot typically be computed analytically or numerically without resorting to solving the corresponding Schr\"odinger or Lippman-Schwinger equation; the Feynman graphs are functions of arbitrary $\mu$ and there is no simplification gained by the expansion in $z$.  

In this Letter we devise a graphical expansion that circumvents these difficulties, and demonstrate its utility by performing analytic calculations of $b_3$ for unitary fermions; it has some features in common with the approach of refs.~  \cite{Bedaque:2002xy,rupak2007universality}. The calculation is mostly analytical, with some integrals performed numerically, and we will show that an extremely accurate determination of $b_3$ can be obtained. The two components of our general procedure are  (i) to perform the ``dual" of the Matsubara sum over discrete frequencies --- in the sense of a Poisson resummation --- which directly leads to a fugacity expansion; (ii) to use a dimer field as in  \cite{Kaplan:1996nv}, which is designed to reproduce the continuum 2-body phaseshift, thereby bypassing discussion of potentials and leading to purely local interactions in space. We consider these two innovations in turn, first addressing the case of free fermions, then including two-body interactions.

\section{Chronographs}

\begin{figure}
\includegraphics[width=6 cm]{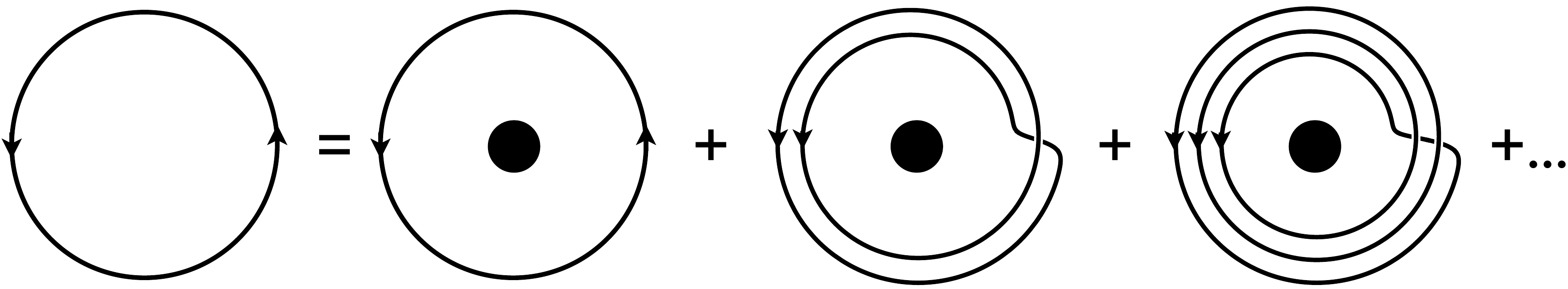}% Here is how to import EPS art
\caption{\label{fig:F1} The free fermion contribution to $  \Omega$. The conventional finite temperature Feynman diagram on the left is expanded as  a sum over worldline loops about the compact time direction (``chronographs") with winding number $\nu$ making a contribution  proportional to $z^\nu$.  The black dot indicates the nontrivial topology, and Euclidian time increases in the counterclockwise direction. }
\end{figure}

We consider a dilute gas  comprised of a single species of nonrelativistic spin half fermion; generalization to bosons or more species is straight forward, but we have not considered the relativistic case.  In the Euclidian time formulation of finite temperature field theory,  $\Omega$ is given by sum of 1PI  vacuum Feynman diagrams, where the theory is analyzed for Euclidian time $\tau$  compactified with period $\beta$ and antiperiodic (periodic) boundary conditions  imposed for fermions (bosons).  
For a free spin $\half$ fermion, $\Omega$ is given by the one loop diagram on the left in Fig.~\ref{fig:F1}.  It is convenient instead to compute  $\partial \Omega/\partial \mu$ with the result
\beq
\frac{\partial\Omega}{\partial\mu} =  -\frac{1}{\beta}\Tr G^E_0=  -\frac{2V}{\beta}\int\frac{d^3\bfp}{(2\pi)^3} \sum_n \widetilde G_0^E(\omega_n,\bfp)
\eeq
where 
$\widetilde G^E_0(\omega, \bfp) =e^{i\omega\,0^+}/(i\omega-(\varepsilon_\bfp -\mu))$ is the free Euclidian propagator, $\varepsilon_\bfp=\bfp^2/2M$, and $\omega_n =2\pi (n+\half)/\beta$; the factor of 2 is from the two spin states, and the minus sign from the fermion loop. The frequency sum is trivial to compute and the result can be subsequently expanded in powers of the fugacity, but it is interesting to note that a Poisson resummation yields the fugacity expansion directly (see also appendix B of Ref.~\cite{Brown:1999xq}):
\beq
\frac{1}{\beta}\sum_n%_{n=-\infty}^\infty 
\widetilde G^E_0(\omega_n,\bfp)& =&\sum_\nu%_{m=-\infty}^\infty 
(-1)^\nu G^E_0(\nu\beta,\bfp)\cr & =& \sum_{\nu=1}^\infty   (-1)^{(\nu+1)} z^\nu e^{-\nu\beta\varepsilon_\bfp } 
\eqn{G0}\eeq
where $G_0^E(\tau,\bfp) =  -\theta(\tau-0^+) e^{-\tau(\varepsilon_\bfp-\mu)}$ is the Fourier transform of $\widetilde G_0^E(\omega,\bfp) $.
The Poisson formula exchanges the sum over Matsubara frequencies for a sum over the winding number $\nu$ of worldlines wrapping around the compact time direction, each term proportional to $z^\nu$, as shown graphically in Fig.~\ref{fig:F1}.  We therefore immediately read off the $b_n$ coefficients for a free fermion:
\beq
b^{(1)}_n = (-1)^{n+1}\frac{ \lambda^3 }{ n}\int\frac{d^3\bfp}{(2\pi)^3}\,e^{- n\beta\varepsilon_\bfp } =\frac{(-1)^{n+1}}{n^{5/2}}\ .
\eqn{ompsi}\eeq
Note this result includes both the $(-1)$ from the Feynman graph, as well as a factor of $(-1)^\nu$ from fermion worldline loops due to antiperiodic boundary conditions.

%%%%%%%%%%%%%%%%%%%%
%%%%%%%%%%%%%%%%%%%%

\begin{figure}[t]
\includegraphics[width=6 cm]{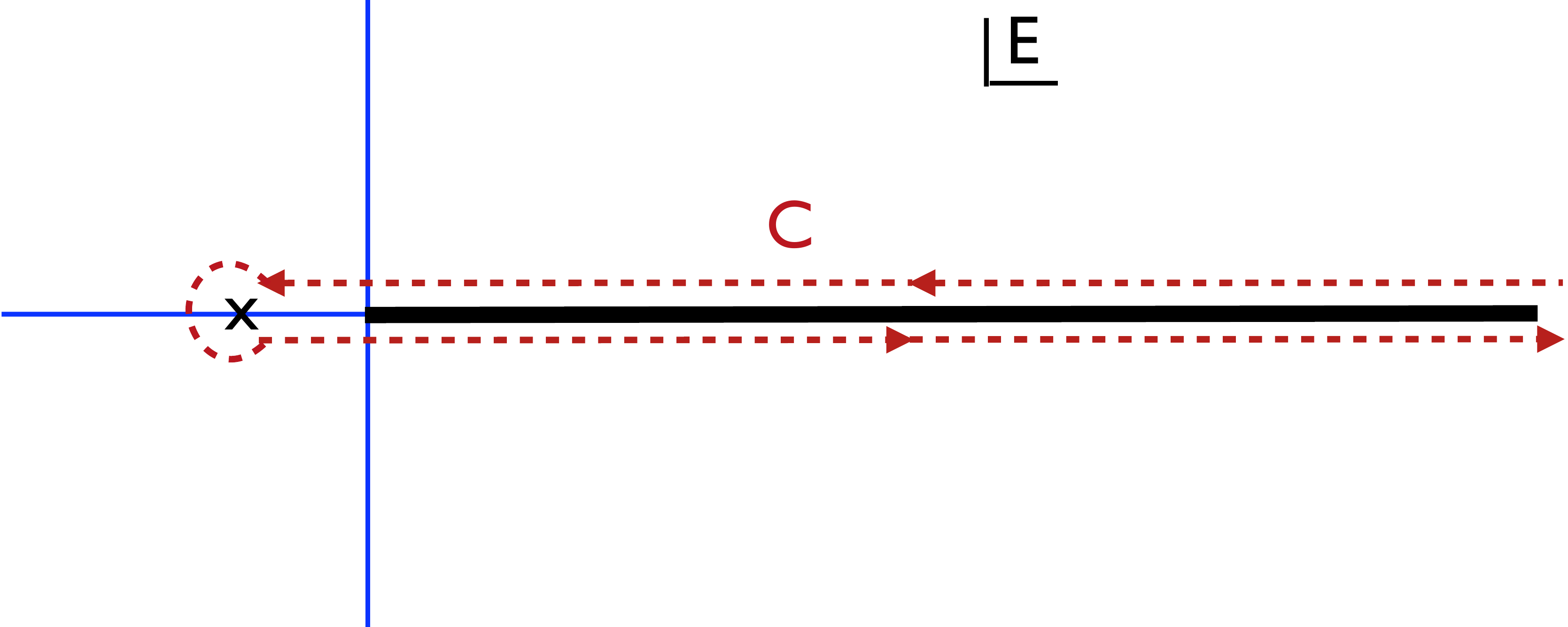}% Here is how to import EPS art
\caption{\label{fig:contour} The contour $C$ in \eq{cont} is designed to pick up contributions from all cuts and poles along the real energy axis. }
\end{figure}

We will refer to the diagrams on the right in Fig.~\ref{fig:F1}  as ``chronographs",
which allow one to compute directly the $n^{th}$ term in the fugacity expansion of $\Omega$ or $\partial\Omega/\partial\mu$.
 The rules for chronographs can be easily generalized for computing the fugacity expansion in interacting systems:
(i) Chronograph propagators $\CG(\tau,\bfp)$  can be defined  in terms  for the Minkowski propagator ${\widetilde G}^M(E,\bfp)=i/(E-\varepsilon_\bfp+i\epsilon)$ at $\mu=0$ via the contour integral along the path $C$  shown in Fig~\ref{fig:contour}, which simply picks up all the physical poles and cuts: 
 \beq
 \CG(\tau,\bfp) \equiv \theta(\tau-0^+)\int_C \frac{dE}{2\pi}\,e^{-E\tau} \widetilde G^M(E,\bfp)
\eqn{cont}
 \eeq
$\CG$ should be thought of as  a multi-valued function of $\tau$ living on a  compact manifold of circumference $\beta$. For a free fermion,
$
\CG_0(\tau,\bfp) = -\theta(\tau) e^{-\tau \varepsilon_\bfp}$;
%\eqn{gprop}
%\eeq
(ii)  vertices (from the Euclidian action) are located on the circle at  Euclidian time $\tau_i$,  each with a 3-momentum conserving $\delta$-function;
(iii)  one integrates over all vertex positions $\tau$ and all propagator 3-momenta $\bfp$;
(iv) a factor of $(-1)^{\nu_F}$ is included where $\nu_F$ is the winding number carried by fermions in the diagram, with an additional $(-1)$ for each closed fermion loop in the parent Feynman diagram;
 (v) symmetry factors are computed as in Feynman diagrams, with the caveat that   two propagators connecting the same two vertices do not warrant a symmetry factor when their length differs by $n\beta$;
 (vi) the winding number about compact Euclidian time, weighted by particle charge, is the order of the graph;  all graphs of order $p$ are included in a fugacity expansion to order $z^p$. For example, a dimer loop with $\nu=1$ contributes to order $z^2$ since the dimer has particle number 2.

%%%%%%%%%%%%%%%%%%%%
%%%%%%%%%%%%%%%%%%%%

\section{Interactions and $b_2$}

To include 2-particle interactions it is convenient to represent the interaction not in terms of a potential,  but  by s-channel dimer exchange, a technique introduced in   \cite{Kaplan:1996nv}.  The advantage is that the dimer --- with  a dispersion relation constructed to exactly reproduce the two particle phase shift $\delta$ ---  has a separable contact interaction with the fermions.  This phase shift is assumed to be given,  either directly from scattering data, or previously calculated from a potential model. That one can take this simplifying approach is due to the fact that the virial coefficients depend on interactions only through the $S$-matrix  \cite{Dashen:1969ep}.

\begin{figure}[t]
\includegraphics[width=8 cm]{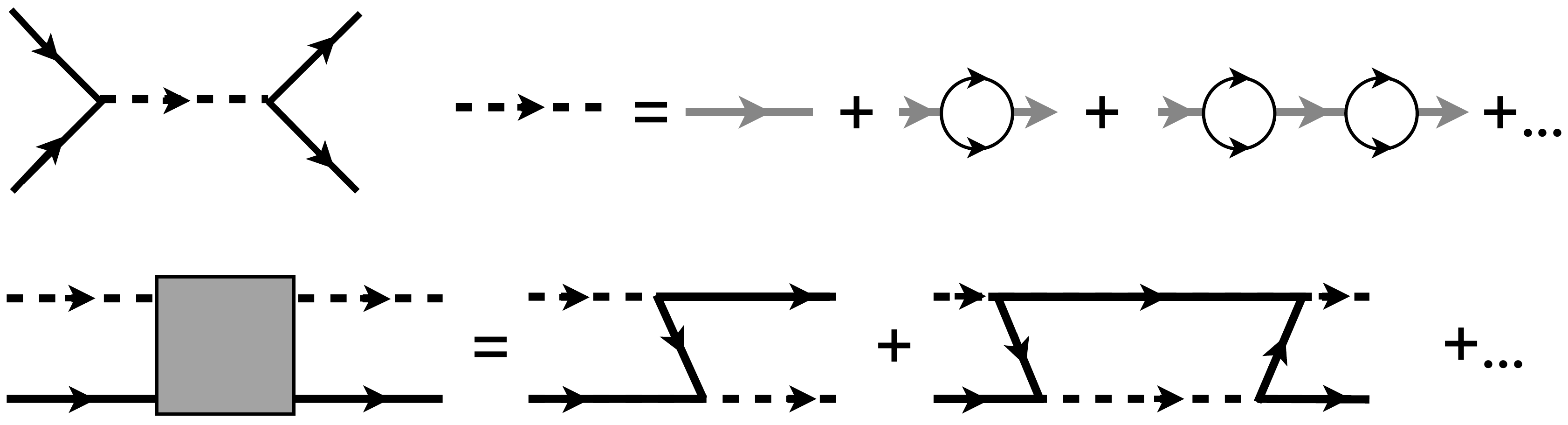}% Here is how to import EPS art
\caption{\label{fig:trans} Feynman graphs for dimer mediated two-body scattering, and the integral equation relevant  for three-body scattering (dashed line = fully dressed dimer, solid line = fermion; gray = $K^{-1}$). }
\end{figure}

We focus on the case of s-wave scattering, commenting below on its generalization to other partial waves.  Consider the Minkowski spacetime Lagrangian
\beq
\CL = \psi^\dagger(i\partial_t +\nabla^2/2M)\psi +\phi^\dagger K \phi +\half \phi^\dagger \psi^T\sigma_2\psi + \text{h.c.}
\eqn{dimerL}\eeq
where $K$ is a function of the Galilean invariant operator $D=(i\partial_t + \nabla^2/4M)$.  
It is important to recognize that this is {\it not} a conventional effective field theory (EFT); in an EFT, one would perform a low energy expansion and express $K$ as a polynomial in $D$, as was done to subleading order in   \cite{Kaplan:1996nv} and to leading order in  \cite{Bedaque:2002xy}; such an expansion of $K$  in powers of $D$ corresponds directly to the effective range expansion of $p\cot\delta(p)$.  However, here we consider $K$ to be a more general function of $D$ (e.g, nonlocal) chosen so that the full dimer Green's function given by the sum in Fig.~\ref{fig:trans}  results in the exact 2-fermion scattering amplitude:
\beq
\widetilde G^M_\phi(E,\bfP)&=& \frac{4\pi}{M} \frac{-i}{k\cot\delta(k) +\sqrt{-k^2}}\ ,
\eqn{gphidef}
\eeq
where $k^2 =[ M(E+2\mu)-\bfP^2/4+ i \epsilon]$,  $E$ and $\bfP$ being the total energy and momentum of the fermion pair.  In   Fig.~\ref{fig:trans} the geometric sum of loop diagrams  creates the correct 2-fermion cut appearing as the $\sqrt{-k^2}$ term in the amplitude; the loops are linearly divergent, and the divergence is absorbed into a constant counterterm in  $K$, so that the renormalized operator is $K_R = K-\text{const.}$ (see, for example  \cite{Kaplan:1998we,Kaplan:1998tg}).  This theory is valid beyond the radius of convergence of the effective range expansion, up to energies where inelastic processes set in, such as pion production in the case where the fermions are nucleons.

%%%%%%%%%%%%%%%
 %%%%%%%%%%%%%%%%%%

 %%%%%%%%%%%%%%%%%%%%%%
\begin{figure}[t]
\includegraphics[width=8.5 cm]{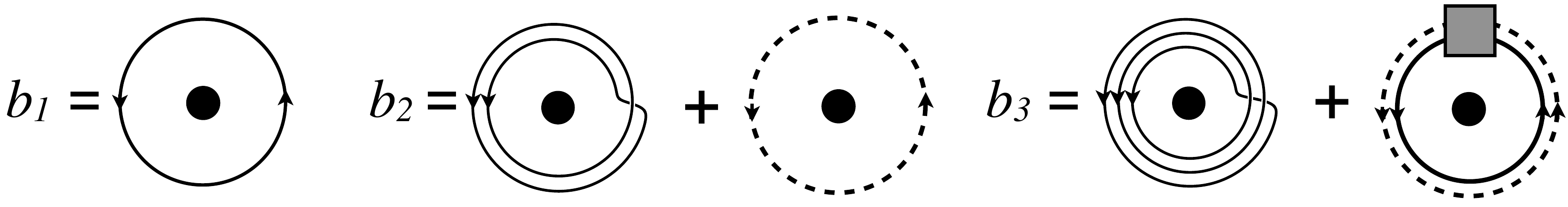}% Here is how to import EPS art
\caption{\label{fig:omega}   Chronograph expansion  for the virial coefficients $b_n$.  Dashed line:  fully dressed dimer propagator,  shaded box:  summed three-body interaction from Fig.~\ref{fig:trans}. For $b_3$, bosons or multiple fermion species would require introduction of a trimer field as well.}
\end{figure}
%%%%%%%%%%%%%%%%%%%%%%%%
The chronographs in this theory for $b_{1,2,3}$ are shown in Fig.~\ref{fig:omega}. The second coefficient $b_2$ gets the contribution $b_2^{(1)}=-1/4\sqrt{2}$ computed in  \eq{ompsi} from free fermions, as well as the dimer contribution computed from $\partial\Omega/\partial\mu$:
\beq
%&b_2^{(2)} =- i\frac{\lambda^3}{2} \int \frac{d^3 \bfP}{(2\pi)^3} \int_C \frac{dE}{2\pi} e^{- \beta E} \ln ({ \widetilde G}^M_\phi)^{-1}\cr
%&=- i\frac{\lambda^3}{2}\int d\mu \int  \frac{d^3\bfP}{(2\pi)^3} \int_C \frac{dE}{2\pi} e^{- \beta E} \left( \widetilde{ G}^M_\phi\frac{\partial ({ \widetilde G}^M_\phi)^{-1}}{\partial k}\frac{ \partial k}{\partial\mu}\right)\notag%\cr&&
b_2^{(2)} =\frac{\lambda^3}{4} \int  \frac{d^3\bfP}{(2\pi)^3} \int_C \frac{dE}{2\pi} e^{- \beta E} \widetilde{ G}^M_\phi\frac{\partial ( i{ \widetilde G}^M_\phi)^{-1}}{\partial \mu}\biggl\vert_{\mu=0}%\cr&&
\eeq
where $ \widetilde{ G}^M_\phi$ and $k$ are given in \eq{gphidef}.  The contour integration picks up contributions both from poles and from  the cut along the positive real $E$ axis from $\sqrt{-k^2}$  in \eq{gphidef}.   If the theory has bound states with binding energy $\varepsilon_n$, then $ \widetilde{G}_\phi^M$ has poles at  $E_n\equiv( \epsilon_n + \bfP^2/4M -2\mu)$ and the integrand in brackets  has poles at $E=E_n$ with residue 2.    To compute the contribution from the cut, one substitutes \eq{gphidef} for $ \widetilde{G}^M_\phi$ accounting for $\sqrt{-k^2}$ flipping sign across the cut.  Combining the cut and pole contributions and performing the  $\bfP$ integration we immediately recover the well-known  result  \cite{Beth:1937zz, Pais:1959zz}
\beq
b^{(2)}_2 = \sqrt{2}\left[\sum_n e^{-\beta\epsilon_n}
 +\frac{1}{\pi} \int_0^\infty dk\, \frac{d\delta(k)}{dk} e^{-\beta \frac{k^2}{M}} \right]\ .
\eeq
This analysis can be extended to other partial waves by introducing new dimer fields with appropriate couplings to fermions.

\section{Computing $b_3$}

% %%%%%%%%%%%%%%%%%%%%%%%%%%%
\begin{figure}[t]
\includegraphics[height=2 cm]{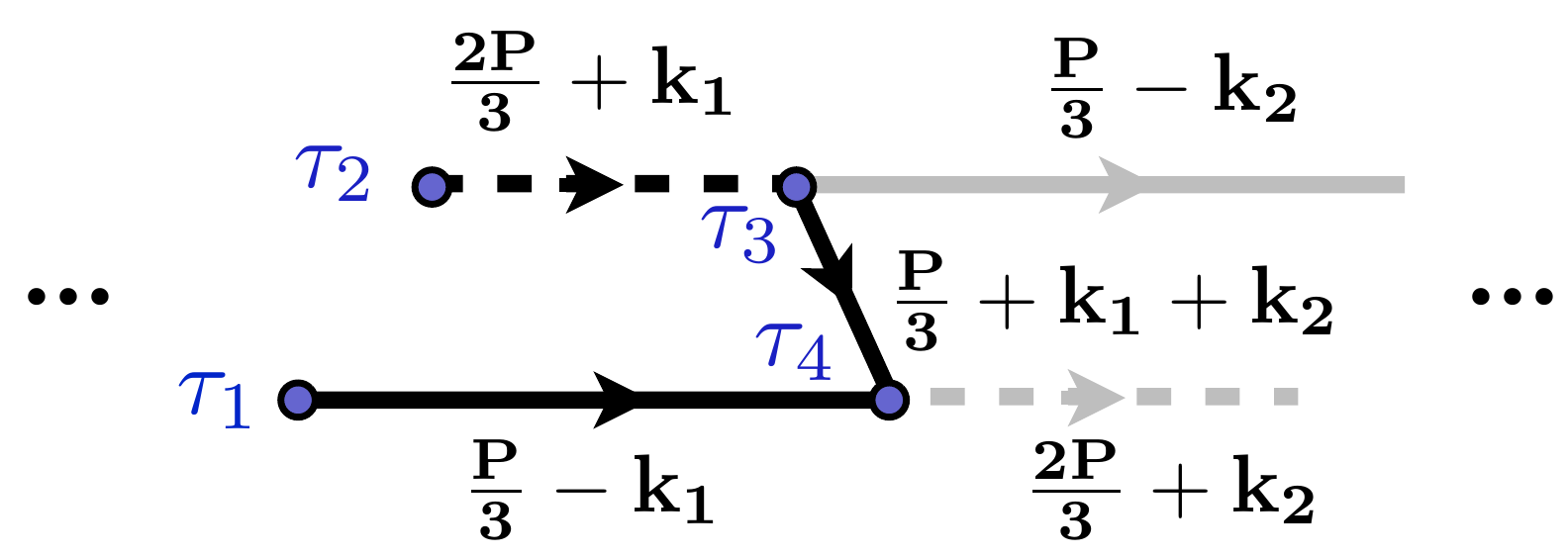}% Here is how to import EPS art
\caption{\label{fig:Ldef}  The three propagator subgraph   (in black) contributing to $b_3$ and corresponding to the expression $L$ appearing in \eqstwo{b3sum}{ldef}.}
\end{figure}
%%%%%%%%%%%%%%%%%%%%%%%%%

At third order we need to compute the new chronographs for $\Omega$    shown in Fig.~\ref{fig:omega}, the first graph yielding the free fermion contribution, $b_3^{(1)}=1/9\sqrt{3}$ as computed in \eq{ompsi}.  The second graph for $b_3^{(3)}$ sums all dimer-fermion and three fermion interactions, barring additional three-body forces; for a single species of fermion renormalizability does not require three-body forces and we shall ignore them here; however, in principle a trimer could be introduced to generate fundamental three-body forces.  This diagram can be expressed in a loop expansion  as
\beq
b_3^{(3)} = \lambda^3 \sum_n \Tr \frac{L^n}{n} 
\eqn{b3sum}\eeq
where $L$ corresponds to the subdiagram in Fig.~\ref{fig:Ldef} with 
\beq
\begin{aligned}
 &L_{12 \bfk_1,34\bfk_2}=\CG_0(\tau_4-\tau_1,\bfP/3-\bfk_1)\cr &\ \ \times \CD(\tau_3-\tau_2,2\bfP/3+\bfk_1)  \CG_0(\tau_4-\tau_3,\bfP/3+\bfk_1+\bfk_2)\ ,
\eqn{ldef}\end{aligned}
\eeq
 $\CG_0$ being the fermion chronograph propagator  and $\CD$ being the dimer propagator computed from \eqstwo{cont}{gphidef}, and we  include a symmetry factor of $1/n$, a spin factor of $\Tr \sigma_2^{2n}=2$ from the vertices and a factor of $(-1)$ due to $\nu_F=3$; there is no overall fermion sign from the parent Feynman diagram.   In this expression, a product of $L$'s corresponds to the integral
\beq
L^2_{12 \bfk_1,56\bfk_2} = \int_0^\beta d\tau_3 d\tau_4 \int \frac{d^3\bfq}{(2\pi)^3}\,L_{12 \bfk_1,34\bfq} L_{34 \bfq,56\bfk_2}
\eeq
with an integral over the center of mass momentum $\bfP$ implied in the trace; this $\bfP$ integral is gaussian and   contributes a factor of $(3\sqrt{3}/\lambda^3)$.  Then the $n^{th}$ term in \eq{b3sum} corresponds to the $(n+1)$-loop contribution to the last diagram in Fig.~\ref{fig:omega}.

%%%%%%%%%%%%%%%%%%%%%%%%%%%
\begin{figure}[t]
\includegraphics[width=6.0 cm]{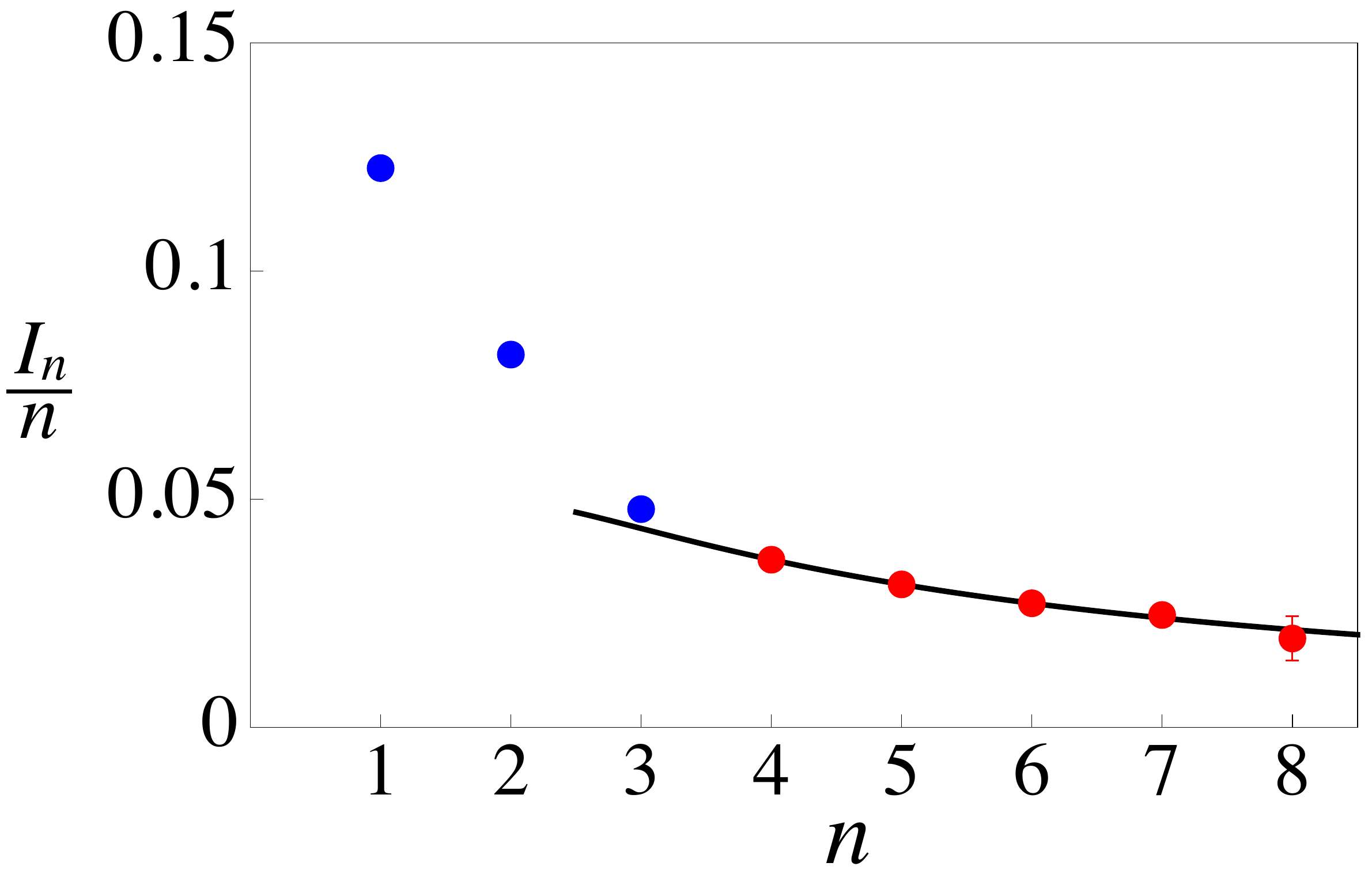}% Here is how to import EPS art
\caption{\label{fig:Ln}  Results for $I_n/n$, including error estimate from numerical integration for $n=3,\ldots,8$.  The solid line is the large-$n$ fit to the last five data points  by the function $(c_0/n+c_1/n^2)$, as described in the text.}
\end{figure}
%%%%%%%%%%%%%%%%%%%%%%%%%

At this point in order to be concrete we will narrow our focus to computing $b^{(3)}_3$ for unitary fermions, for which $ p\cot\delta=0 $ identically.  From \eqstwo{cont}{gphidef} we find the dimer propagator
$
\CD(\tau,\bfq) =- \theta(\tau) \sqrt{16\pi/M^{3}}\,e^{-\tau \bfq^2/4M}/\sqrt{\tau}$.
In this case  the expansion takes the form 
\beq
b_3^{(3)}=3^{3/2}\sum_{n=1}^\infty\frac{(-1)^n  I_n}{n}
\eeq
where $I_n$ is given by the integral
\beq
\begin{aligned}
&\left(16\pi \right)^{n/2} %\cr &\times
 \int_\CT \prod_{i=1}^{n} \frac{d\tau_{2i-1}d\tau_{2i}}{\sqrt{\tau_{2i+1}-\tau_{2i}}}
\int\prod_{j=1}^n \frac{d^3\bfk_j}{(2\pi)^3} e^{-\bfk_a\cdot A_{ab}\bfk_b}\cr
&={\left(2\pi \right)^{-n}}  \int_\CT \prod_{i=1}^{n} \frac{d\tau_{2i-1}d\tau_{2i}}{\sqrt{\tau_{2i+1}-\tau_{2i}}}
\left(\det A\right)^{-\frac{3}{2}} 
\end{aligned}
\eqn{b33}\eeq
 which is positive and independent of $\beta$ and $M$,  $A$ being the $n\times n$ matrix
\beq
\begin{aligned}
A_{ab} &= \delta_{ab}\frac{4\tau_{2a+2}-\tau_{2a+1} + \tau_{2a}-4\tau_{2a-1}  }{4}\cr
&+\left(\hat\delta_{a,b+1}\frac{\tau_{2a}-\tau_{2a-1}}{2}+a \leftrightarrow b
\right)
\end{aligned}\eeq
with $\hat\delta_{ab}$ is the Kronecker $\delta$-function with indices defined modulo $n$, so that $\hat \delta_{n+a,b}=\hat\delta_{a,n+b} = \delta_{ab}$ for $1\le a,b\le n$, and  
$\int_\CT$ represents a $2n$-dimensional time ordered integral over $\tau_i$  with $0\le \tau_1\le \tau_2<\ldots <\tau_{2n}\le  1$ and $\tau_{2n+j} \equiv 1+\tau_j$ for $j=1,2$.
 
The integral $ I_n$ can be performed analytically for $n=1,2$ with the result $ I_1 = 2/(3\sqrt{3}\pi)$, $ I_2 = 8/(9\sqrt{3}\pi)$; for $n=3,\ldots,8$ we have computed the integrals numerically.
We find that $I_n$  is apparently a smooth function of $n$ for large $n$ and we perform a large-$n$ extrapolation, fitting our results for $n=4,\ldots,8$ to the function $I_n \sim (c_0+c_1/n)$ finding $c_0=\czval\pm\czerr$, $c_1=\coval\pm\coerr$ (with highly correlated errors);  the results are shown in Fig.~\ref{fig:Ln}.  We estimate our errors by also fitting to the function  $(c_0+c_1/n + c_2/n^2)$, and varying the range of points used in the fit, finding a very stable result.  Our final result for the interacting contribution to $b_3$ is $b_3^{(3)} =\bttval\pm\btterr$, or for the full answer, $b_3 =\btval\pm \bterr$, to be compared with the recent computation  $b_3 = -0.29095295$ by Liu, Hu and Drummond  \cite{liu2009virial}, which involved summing over  energy levels for three unitary fermions in a harmonic trap \cite{Werner:2006zz}.
%Note that our fit for $c_0$ implies that in the limit of an infinite number of loops,  $\lim_{n\to\infty} I_n =c_0$ is a finite number; a more detailed %investigation will appear elsewhere. 
 It is remarkable that an expansion and extrapolation of chronographs is able to arrive at such a precise number for what is essentially a nonperturbative system; however,  we do not have an explanation for why our procedure leads to an estimated error for $b_3$ ($\pm\bterr$)  which is significantly smaller than our discrepancy with Liu et al. ($\discrep$).

To our knowledge, this is the first time $b_3$ was calculated by analytical means for a strongly interacting system, and it suggests that the graphical techniques presented here for the virial expansion may prove powerful for applications to other systems as well.
 \medskip
 
 \begin{acknowledgments}
We thank   L. Brown, M. Savage, and  D. Son for helpful conversations.  This work was supported in part by U.S.\ DOE grant No.\ DE-FG02-00ER41132.
\end{acknowledgments}
\bibliography{virial}

\begin{thebibliography}{13}
\expandafter\ifx\csname natexlab\endcsname\relax\def\natexlab#1{#1}\fi
\expandafter\ifx\csname bibnamefont\endcsname\relax
  \def\bibnamefont#1{#1}\fi
\expandafter\ifx\csname bibfnamefont\endcsname\relax
  \def\bibfnamefont#1{#1}\fi
\expandafter\ifx\csname citenamefont\endcsname\relax
  \def\citenamefont#1{#1}\fi
\expandafter\ifx\csname url\endcsname\relax
  \def\url#1{\texttt{#1}}\fi
\expandafter\ifx\csname urlprefix\endcsname\relax\def\urlprefix{URL }\fi
\providecommand{\bibinfo}[2]{#2}
\providecommand{\eprint}[2][]{\url{#2}}

\bibitem[{\citenamefont{Liu et~al.}(2009)\citenamefont{Liu, Hu, and
  Drummond}}]{liu2009virial}
\bibinfo{author}{\bibfnamefont{X.}~\bibnamefont{Liu}},
  \bibinfo{author}{\bibfnamefont{H.}~\bibnamefont{Hu}}, \bibnamefont{and}
  \bibinfo{author}{\bibfnamefont{P.}~\bibnamefont{Drummond}},
  \bibinfo{journal}{Phys. Rev. Lett.} \textbf{\bibinfo{volume}{102}},
  \bibinfo{pages}{160401} (\bibinfo{year}{2009}), ISSN
  \bibinfo{issn}{1079-7114}.

\bibitem[{\citenamefont{Kaplan et~al.}(1998{\natexlab{a}})\citenamefont{Kaplan,
  Savage, and Wise}}]{Kaplan:1998tg}
\bibinfo{author}{\bibfnamefont{D.~B.} \bibnamefont{Kaplan}},
  \bibinfo{author}{\bibfnamefont{M.~J.} \bibnamefont{Savage}},
  \bibnamefont{and} \bibinfo{author}{\bibfnamefont{M.~B.} \bibnamefont{Wise}},
  \bibinfo{journal}{Phys.Lett.} \textbf{\bibinfo{volume}{B424}},
  \bibinfo{pages}{390} (\bibinfo{year}{1998}{\natexlab{a}}).

\bibitem[{\citenamefont{Kaplan et~al.}(1998{\natexlab{b}})\citenamefont{Kaplan,
  Savage, and Wise}}]{Kaplan:1998we}
\bibinfo{author}{\bibfnamefont{D.~B.} \bibnamefont{Kaplan}},
  \bibinfo{author}{\bibfnamefont{M.~J.} \bibnamefont{Savage}},
  \bibnamefont{and} \bibinfo{author}{\bibfnamefont{M.~B.} \bibnamefont{Wise}},
  \bibinfo{journal}{Nucl.Phys.} \textbf{\bibinfo{volume}{B534}},
  \bibinfo{pages}{329} (\bibinfo{year}{1998}{\natexlab{b}}).

\bibitem[{\citenamefont{Bedaque and Rupak}(2003)}]{Bedaque:2002xy}
\bibinfo{author}{\bibfnamefont{P.~F.} \bibnamefont{Bedaque}} \bibnamefont{and}
  \bibinfo{author}{\bibfnamefont{G.}~\bibnamefont{Rupak}},
  \bibinfo{journal}{Phys.Rev.} \textbf{\bibinfo{volume}{B67}},
  \bibinfo{pages}{174513} (\bibinfo{year}{2003}).

\bibitem[{\citenamefont{Rupak}(2007)}]{rupak2007universality}
\bibinfo{author}{\bibfnamefont{G.}~\bibnamefont{Rupak}},
  \bibinfo{journal}{Phys. Rev. Lett.} \textbf{\bibinfo{volume}{98}},
  \bibinfo{pages}{90403} (\bibinfo{year}{2007}).

\bibitem[{\citenamefont{Luo et~al.}(2007)}]{PhysRevLett.98.080402}
\bibinfo{author}{\bibfnamefont{L.}~\bibnamefont{Luo}} \bibnamefont{et~al.},
  \bibinfo{journal}{Phys. Rev. Lett.} \textbf{\bibinfo{volume}{98}},
  \bibinfo{pages}{080402} (\bibinfo{year}{2007}).

\bibitem[{\citenamefont{{Nascimb{\`e}ne} et~al.}(2010)}]{2010Natur.463.1057N}
\bibinfo{author}{\bibfnamefont{S.}~\bibnamefont{{Nascimb{\`e}ne}}}
  \bibnamefont{et~al.}, \bibinfo{journal}{\nat} \textbf{\bibinfo{volume}{463}},
  \bibinfo{pages}{1057} (\bibinfo{year}{2010}), \eprint{0911.0747}.

\bibitem[{\citenamefont{Kaplan}(1997)}]{Kaplan:1996nv}
\bibinfo{author}{\bibfnamefont{D.~B.} \bibnamefont{Kaplan}},
  \bibinfo{journal}{Nucl.Phys.} \textbf{\bibinfo{volume}{B494}},
  \bibinfo{pages}{471} (\bibinfo{year}{1997}).

\bibitem[{\citenamefont{Brown and Yaffe}(2001)}]{Brown:1999xq}
\bibinfo{author}{\bibfnamefont{L.~S.} \bibnamefont{Brown}} \bibnamefont{and}
  \bibinfo{author}{\bibfnamefont{L.~G.} \bibnamefont{Yaffe}},
  \bibinfo{journal}{Phys.Rept.} \textbf{\bibinfo{volume}{340}},
  \bibinfo{pages}{1} (\bibinfo{year}{2001}), \eprint{physics/9911055}.

\bibitem[{\citenamefont{Dashen et~al.}(1969)\citenamefont{Dashen, Ma, and
  Bernstein}}]{Dashen:1969ep}
\bibinfo{author}{\bibfnamefont{R.}~\bibnamefont{Dashen}},
  \bibinfo{author}{\bibfnamefont{S.-K.} \bibnamefont{Ma}}, \bibnamefont{and}
  \bibinfo{author}{\bibfnamefont{H.~J.} \bibnamefont{Bernstein}},
  \bibinfo{journal}{Phys.Rev.} \textbf{\bibinfo{volume}{187}},
  \bibinfo{pages}{345} (\bibinfo{year}{1969}).

\bibitem[{\citenamefont{Beth and Uhlenbeck}(1937)}]{Beth:1937zz}
\bibinfo{author}{\bibfnamefont{E.}~\bibnamefont{Beth}} \bibnamefont{and}
  \bibinfo{author}{\bibfnamefont{G.}~\bibnamefont{Uhlenbeck}},
  \bibinfo{journal}{Physica} \textbf{\bibinfo{volume}{4}}, \bibinfo{pages}{915}
  (\bibinfo{year}{1937}).

\bibitem[{\citenamefont{Pais and Uhlenbeck}(1959)}]{Pais:1959zz}
\bibinfo{author}{\bibfnamefont{A.}~\bibnamefont{Pais}} \bibnamefont{and}
  \bibinfo{author}{\bibfnamefont{G.}~\bibnamefont{Uhlenbeck}},
  \bibinfo{journal}{Phys.Rev.} \textbf{\bibinfo{volume}{116}},
  \bibinfo{pages}{250} (\bibinfo{year}{1959}).

\bibitem[{\citenamefont{Werner and Castin}(2006)}]{Werner:2006zz}
\bibinfo{author}{\bibfnamefont{F.}~\bibnamefont{Werner}} \bibnamefont{and}
  \bibinfo{author}{\bibfnamefont{Y.}~\bibnamefont{Castin}},
  \bibinfo{journal}{Phys.Rev.Lett.} \textbf{\bibinfo{volume}{97}},
  \bibinfo{pages}{150401} (\bibinfo{year}{2006}).

\end{thebibliography}
\end{document}